\documentclass[5p,times,sort&compress]{elsarticle}

\pdfoutput=1
\usepackage{siunitx}
\usepackage[caption=false]{subfig}

\journal{Nuclear Instruments and Methods in Physics Research A}

\usepackage[pscoord]{eso-pic}
\newcommand{\placetextbox}[3]{
\setbox0=\hbox{#3}
\AddToShipoutPictureFG*{ \put(\LenToUnit{#1\paperwidth},\LenToUnit{#2\paperheight}){\vtop{{\null}\makebox[0pt][c]{#3}}}
}
}
\placetextbox{.5}{0.055}{~\copyright~2016. This manuscript version is made available under the CC-BY-NC-ND 4.0 license http://creativecommons.org/licenses/by-nc-nd/4.0/}

\begin{document}

\begin{frontmatter}

\title{Using the Automata Processor for Fast Pattern Recognition in High Energy Physics Experiments - A Proof of Concept}

\author[addrfnal]{Michael H. L. S. Wang\corref{mwcorrauthor}}
\cortext[mwcorrauthor]{Corresponding author}
\ead{mwang@fnal.gov}

\author[addrfnal]{Gustavo Cancelo}
\author[addrfnal]{Christopher Green}
\author[addruva]{Deyuan Guo}
\author[addruva]{Ke Wang}
\author[addrfnal]{Ted Zmuda}

\address[addrfnal]{Fermi National Accelerator Laboratory, Batavia, IL 60510, USA}
\address[addruva]{University of Virginia, Charlottesville, VA 22904, USA}

\begin{abstract}
We explore the Micron Automata Processor (AP) as a suitable commodity technology that can address the growing computational needs of pattern recognition in High Energy Physics (HEP) experiments.  A toy detector model is developed for which an electron track confirmation trigger based on the Micron AP serves as a test case. Although primarily meant for high speed text-based searches, we demonstrate a proof of concept for the use of the Micron AP in a HEP trigger application.\end{abstract}


\end{frontmatter}

\section{Introduction}

Pattern recognition occupies a central role in the reconstruction and analysis chains of practically all High Energy Physics (HEP) experiments. The ability to recognize a charged particle track from a pattern of detector ``hits",  for example, is extremely important because the trajectory of a particle carries crucial information on its properties and provides a powerful signature to isolate it from unwanted background.  With the advent of electronic detectors and readout systems nearly 50 years ago~\cite{ref:charpak}, tasks, like the manual scanning of tracks, were transformed into computational problems. Such pattern recognition problems have grown more challenging with every new generation of experiment due to the trend towards more complex event topologies and higher particle densities.  To cope with this trend, offline reconstruction applications have so far relied on the rough doubling of transistors in many-core architectures every two years (Moore's Law), while online applications with real-time requirements have traditionally relied on custom hardware solutions.  Unfortunately, Moore's Law becomes less dependable as we enter a period of diminishing performance returns where power dissipation issues from leakage currents dominate as we approach the atomic scale~\cite{ref:leakage}.  On the other hand, custom hardware solutions often entail more technical risks and require more investments in manpower and capital.

In this paper, we take a different approach by exploring emerging commercial technologies designed to deal with the deluge of digital information in today's data centered economies.  One such technology is the Micron Automata Processor (AP)~\cite{ref:ieeeap} which is specifically targeted at pattern matching applications like those in the Internet search industry and bioinformatics~\cite{ref:armuva,ref:brilluva,ref:motif}.  It is a direct hardware implementation of Non-deterministic Finite Automata (NFA) and can simultaneously apply thousands of rules to find patterns in data streams at a constant rate of 1 Gbps/chip.  As a proof of concept to demonstrate its feasibility for HEP pattern finding applications, we develop a simple toy detector model typical of those found at modern hadron colliders and use the Micron AP to implement an electron track confirmation trigger.

\section{The Automata Processor}

\subsection{Hardware Architecture}

Because the Micron AP is based on a new and radical, non-von Neumann architecture, this section will provide a description of the hardware, focusing on aspects relevant to our evaluation.

\subsubsection{A Memory-based Design}
\begin{figure}[htbp]
\centering
\includegraphics[width=0.45\textwidth]{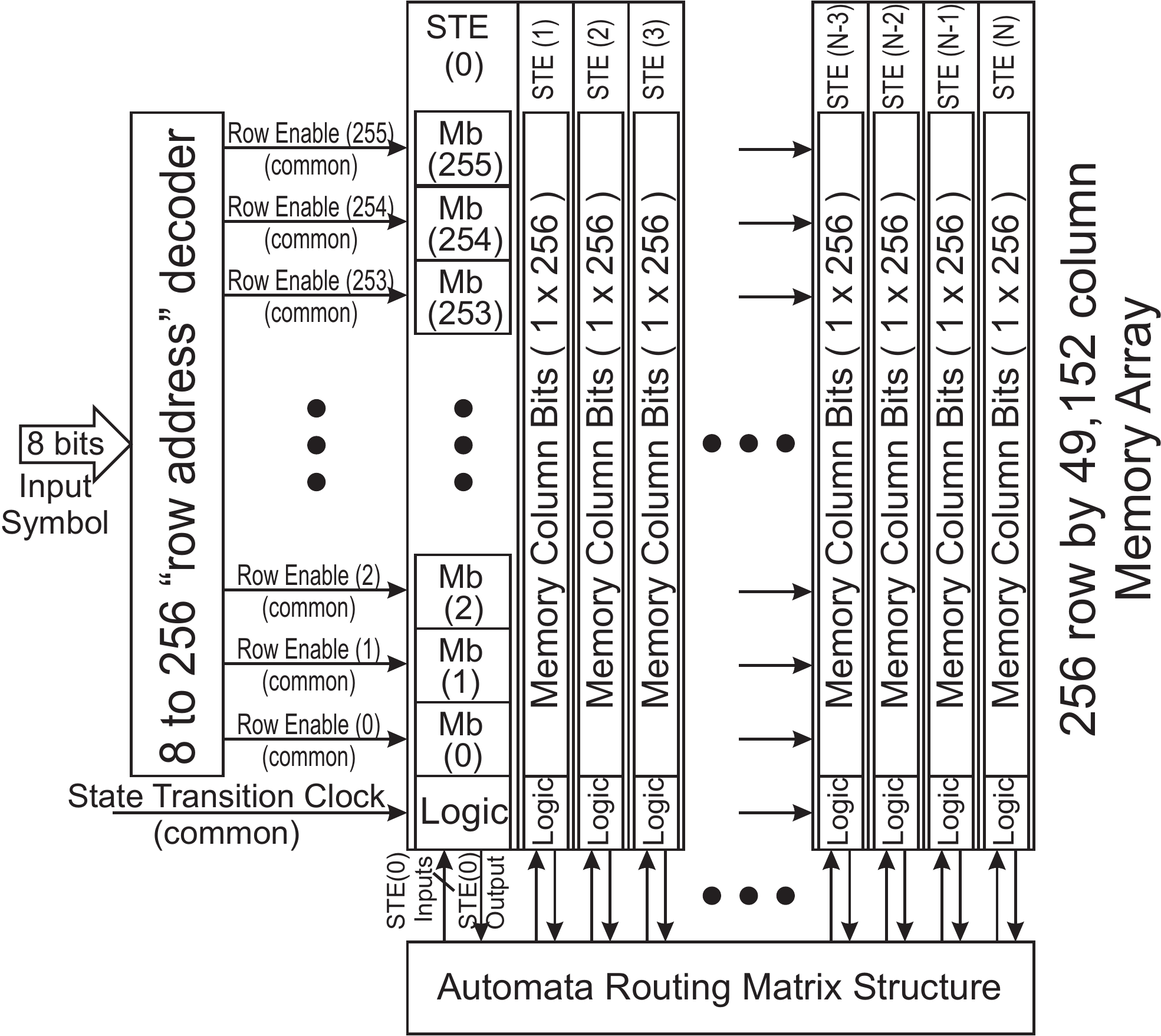}
\caption{The 2D memory array architecture of the Micron Automata Processor adapted from conventional SDRAM technology.}
\label{fig:blockhw1}
\end{figure}

The Micron AP is derived from conventional SDRAM technology and its hardware architecture (see Figure~\ref{fig:blockhw1}) is conveniently understood in terms of a two-dimensional memory array.  Each input byte on an 8-bit wide bus serves as a \emph{row address} which is presented to an 8-to-$2^8$ decoder that selects one out of 256 possible rows.  A column of 256 cells in this array, together with additional logic for state information, comprises the basic building block of the AP known as a State Transition Element (STE).  Any cell or combination of cells in an STE can be programmed to recognize any subset of $2^8$ possible values or \emph{symbols}.  When the address decoder selects a cell in an enabled column or STE, which is programmed to recognize its associated symbol, the stored value of ``1" is output to indicate symbol recognition.  This causes the STE output to change state and can be used to enable other downstream STEs. Multiple STEs, each programmed to recognize specific symbols, can be chained together to recognize patterns or strings of symbols.  Such patterns need not be limited to ASCII strings and could just as well represent hit addresses associated with a particle trajectory in a tracking detector.  The interconnections between the STEs is provided by the routing matrix structure, represented by the block at the bottom of Figure~\ref{fig:blockhw1}, which plays the role of the column address and decode operations.

\subsubsection{Reporting Pattern Matches}\label{sec:apreparch}
The Micron AP was specifically designed to perform high-speed pattern recognition.  To be useful in this application, there must be an efficient way to tell whether pattern matches were found and provide details on these matches.   To this end, any AP building block such as an STE, counter, or boolean can be configured to generate a signal known as a \emph{report event} whenever it recognizes an input symbol.  This way,  the last element in  a string of STEs programmed to recognize a sequence of symbols in an expression, can generate such a signal to indicate a matching pattern in the input data stream.

\begin{figure}[htbp]
\centering
\includegraphics[width=0.45\textwidth]{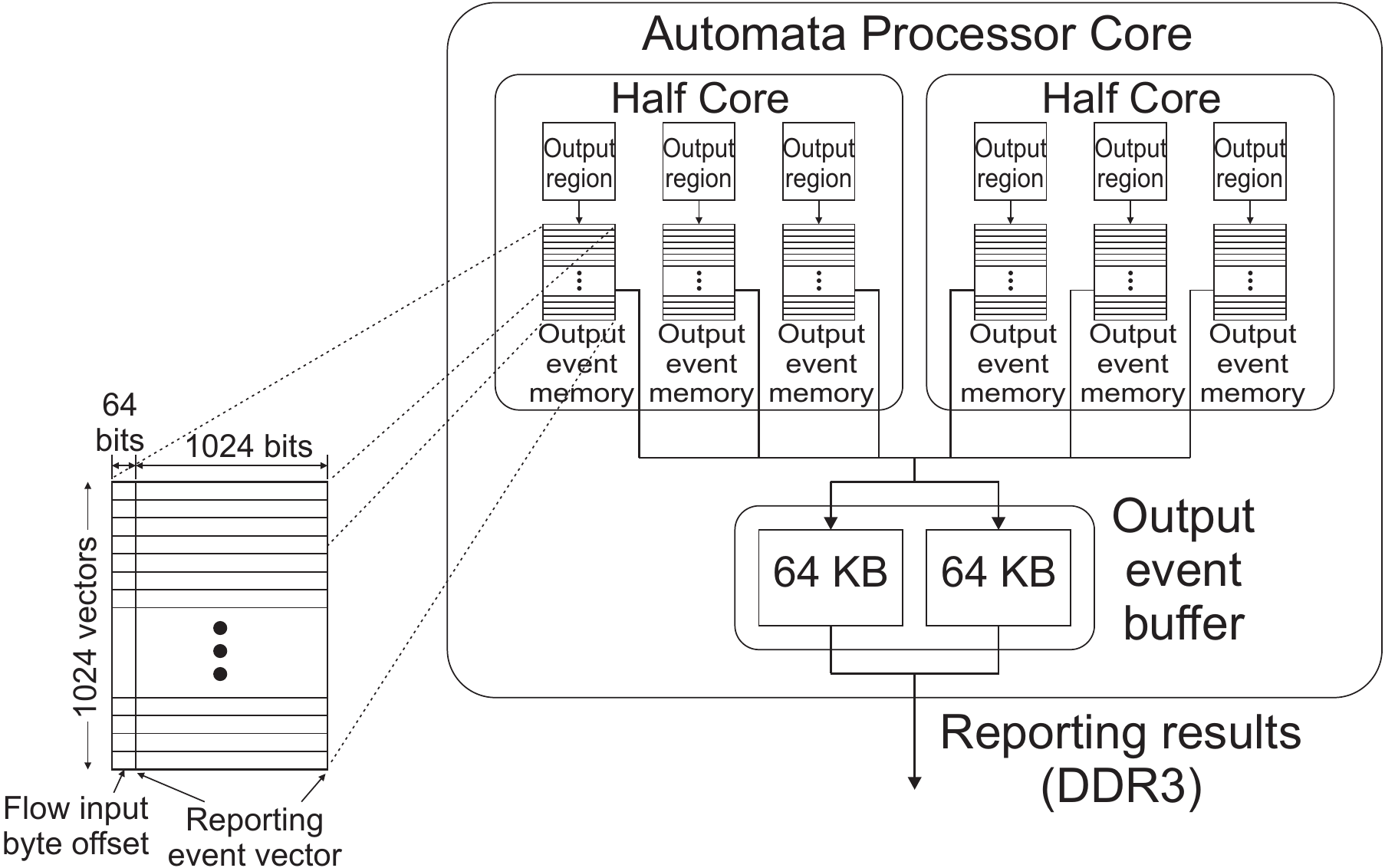}
\caption{Reporting event vector readout architecture.}
\label{fig:blockevtvec}
\end{figure}

The portion of the AP architecture relevant to match reporting is shown in Figure~\ref{fig:blockevtvec}.   The AP is divided into 2 half cores, each of which has 3 output regions.  Each region has a local storage area known as the \emph{output event memory} with room for 1024 \emph{output event vectors} (or \emph{report vectors}) that are up to 1088 bits wide.  Each vector has a 64-bit preamble used to indicate the offset from the start of the input data stream.  The remaining 1024 bits are used to identify the element in the region that generated the report event.  In this respect, the AP is unique because it provides both spatial and temporal information on the matches.  The maximum number of reporting events per symbol cycle is therefore $2\:cores\times3\:regions/core\times1024\:events/region/cycle=6144\:events/cycle$.  The compilation process will fail if the total number of reporting elements exceeds this number.

An output event vector is generated on every symbol cycle for which there was a report event in that region.  Since each vector is associated with a symbol cycle, events occurring on the same cycle require a single vector while events on different cycles require separate vectors.  When symbol processing completes, the vectors are transferred from each region's output event memory to the output event buffer, where they can be read by external hardware.

\subsubsection{Additional Latencies}
Ideally, the total processing time would depend only on the number of input symbols.  In reality, there are overheads tied to the internal memory transfers described above which introduce additional latencies. There is an overhead associated with transferring each vector from a region's event memory to the output buffer.  There is also a start-up overhead associated with transferring the first vector to the output buffer.  Just to determine that a region is empty also incurs an overhead.  The overhead due to transferring each vector depends on its size and can be reduced by configuring the AP  to use smaller vectors.

\subsection{Programming the Automata Processor}
The Micron AP is programmed using tools provided in the AP Software Development Kit (SDK).  The process begins by creating a human-readable representation of the automata which can be done through a graphical tool known as the AP Workbench.  This representation is transformed into machine-readable form in the compilation step which involves optimization and placing and routing the automata elements onto the AP fabric.  This produces a binary relocatable image that is loaded and executed on the AP hardware.

\section {Proof of Principle: A Pixel-augmented Electron Confirmation Trigger}\label{sec:poc}

\begin{table*}[t]
\centering
\begin{center}
\begin{tabular}{S[table-format=1]S[table-format=2.] S[table-format=2] S[table-format=3] S[table-format=4]S[table-format=5] S[table-format=4] S[table-format=8]}
\hline \hline
{Layer} & {Radius} & {Faces}  & {Modules} & {ROCs} & {Pixels}  & {Pixels} & {Total pixels} \\
&{(cm)}&&&&{($\phi$)}&{(z)}& \\
\hline
1 &   2.99 & 12 &    96 & 1536 & 1920   & 3328  & 6389760 \\
2 &   6.99 &  28 & 224 & 3584 & 4480   & 3328  & 14909440 \\
3 & 10.98   &  44 & 352 & 5632 & 7040   & 3328  & 23429120 \\
4 & 15.97   &  64 & 512 & 8192 & 10240 & 3328  & 34078720 \\
\hline
\multicolumn{5}{c}{Length of the toy pixel detector} & \multicolumn{3}{c}{54.88 cm} \\
\hline
\end{tabular}
\end{center}
\caption{Toy pixel detector specifications.  Each face is a ladder consisting of 8 sensor modules.  Each sensor module is made up of a pixel sensor bonded to 2 rows of 8 Read-Out-Chips (ROCs) each. Each ROC has $80$ rows $\times$ $52$ columns of pixels. Each pixel sensor measures $165\times98 \mu$m.}
\label{tab:bpixtoy}
\end{table*}

\begin{figure}[htbp]
\centering
\includegraphics[width=0.45\textwidth]{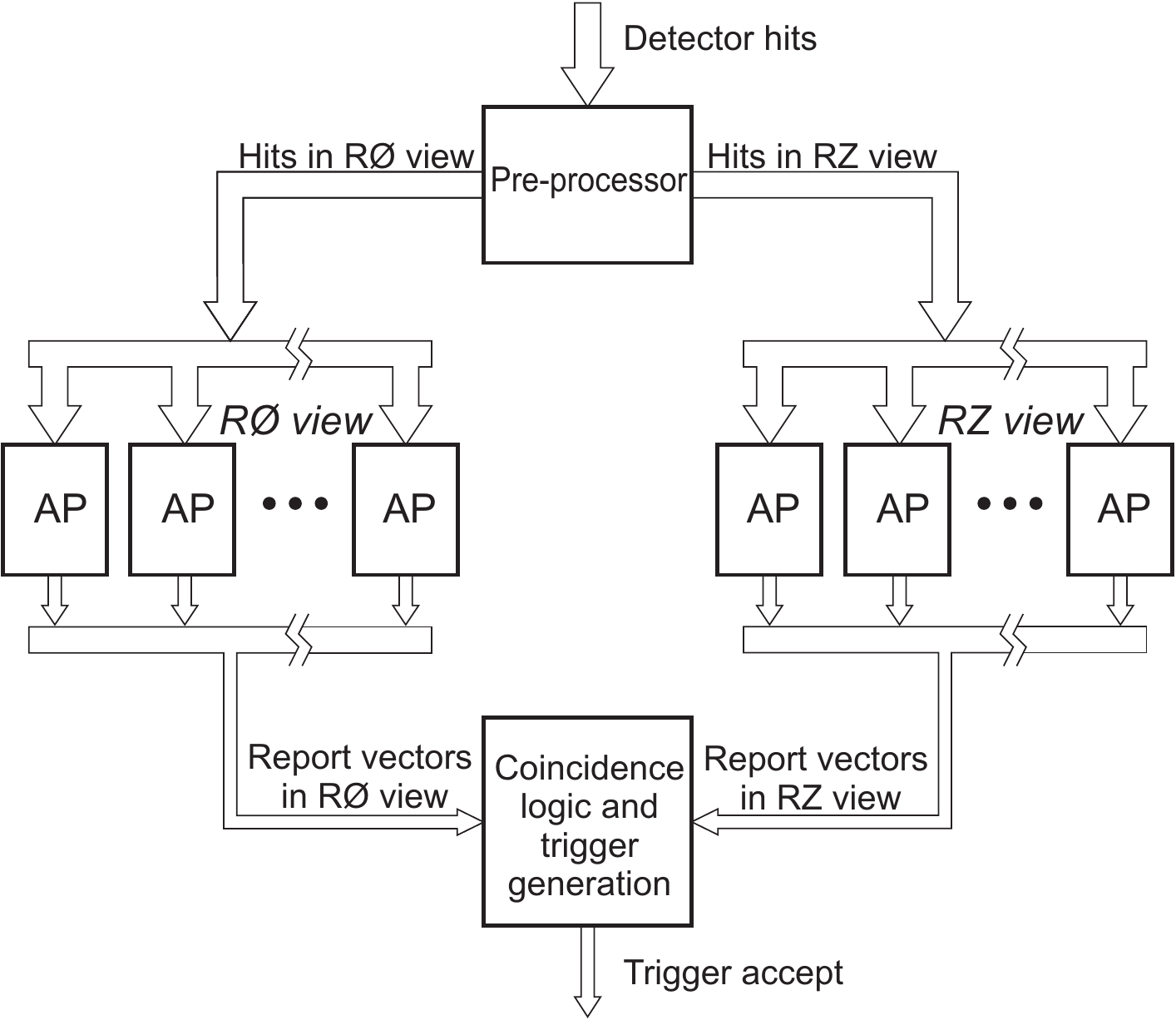}
\caption{Block diagram of Automata-Processor-based electron track confirmation trigger hardware.}
\label{fig:aptrigblock}
\end{figure}

To investigate the feasibility of the Micron AP for HEP pattern recognition applications, we consider an electron confirmation trigger application where isolated high $p_T$ electrons are verified and confirmed in a hadron collider detector by matching energy clusters in an electromagnetic (EM) calorimeter with charged particle tracks in a pixel-based tracking detector. A simplified block diagram depicting the trigger hardware architecture is shown in Figure~\ref{fig:aptrigblock}.  Incoming pixel detector hit data are decomposed into the $R-\phi$ (\emph{bend} plane) and $R-Z$ (\emph{non-bend} plane) views and fed into two separate banks of AP chips that perform the electron track confirmation in each view~\cite{ref:3dpats}.  The reporting event vectors in each view are read out of each bank and fed into external logic that determines if there is at least one pair of reporting events (one from each view) that are correlated in time.  If this condition is satisfied,  a \emph{trigger accept} signal is generated.

\subsection{Toy Detector}\label{sec:bpixtoy}

\begin{figure}[htbp]
\centering
\includegraphics[width=0.45\textwidth]{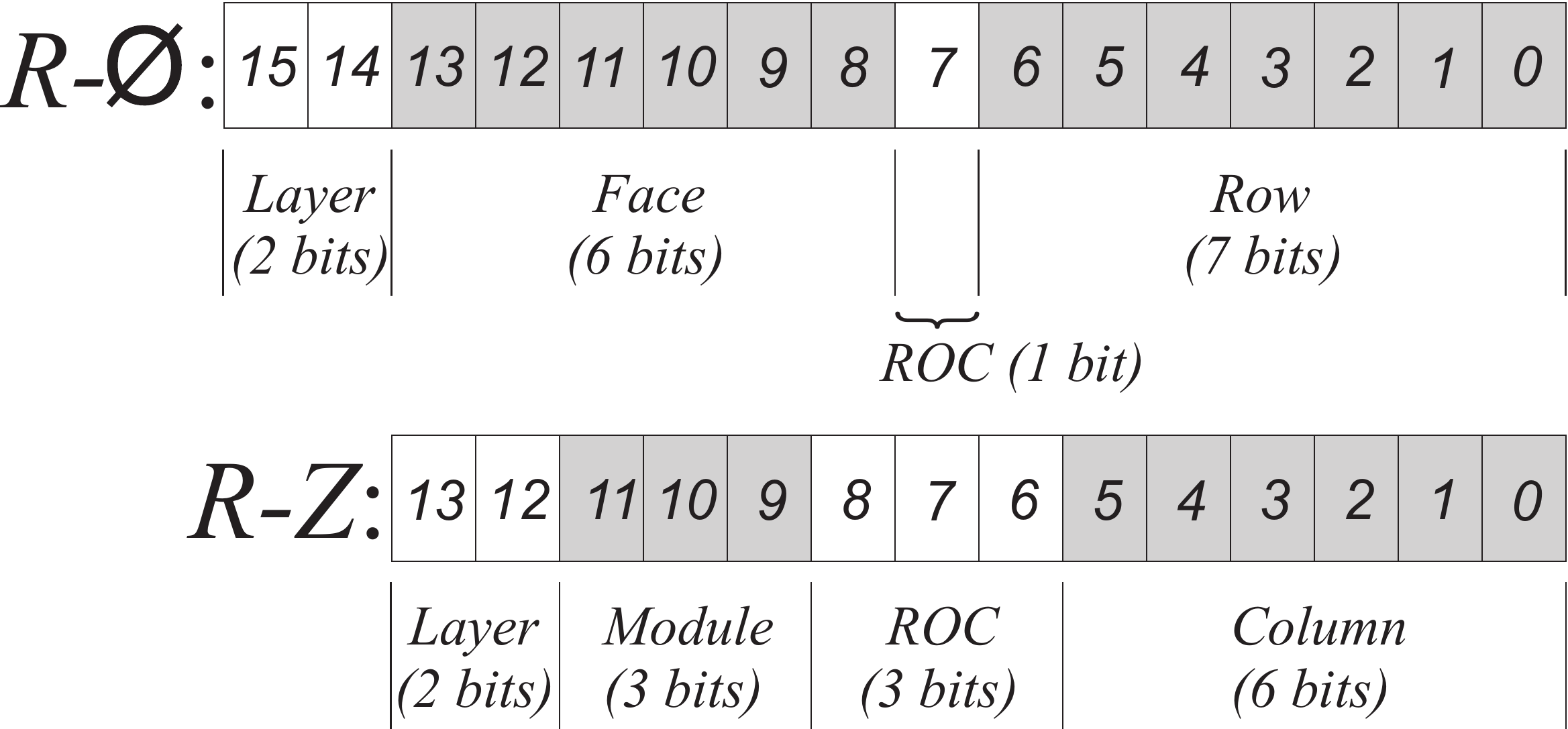}
\caption{Data encoding scheme for pixel addresses in the $R-\phi$ and $R-z$ views.}
\label{fig:bitencoding}
\end{figure}

For our proof-of-principle studies, we developed a toy detector model consisting of 4 concentric cylinders (layers) approximating the barrel portion of the CMS Phase-1 pixel detector~\cite{ref:CMSpixuptdr}.  The geometry and specifications of this model are described in Table~\ref{tab:bpixtoy}. All pixels in the entire detector have dimensions measuring $165\times98$ $\mu$m with the longer axis oriented along the beam axis ($z$).   The pixels in each layer form a uniform grid laid out over the entire cylindrical surface.  This means that the face of each pixel is really a cylindrical tube segment instead of a flat rectangular area.  Each of the 4 detector layers are uniformly divided in azimuthal angle into faces or \emph{ladders} that are parallel to the $z$-axis and run the entire length of the cylinder.  Each ladder is divided in $z$ into 8 equal sections or \emph{modules}.  Each module is further subdivided into 2 rows (along azimuthal direction) by 8 columns (along $z$) of Read-Out-Chips (ROCs).  Each ROC, in turn, consists of 80 rows by 52 columns of pixels. Pixels are uniquely identified using a cylindrical coordinate system ($R$, $\phi$, $z$), consisting of two orthogonal views which are projections onto the $R-\phi$ and $R-z$ planes. Pixel addresses or coordinates are encoded using the 16-bit  and 14-bit data formats shown in Figure~\ref{fig:bitencoding} for the $R-\phi$ and $R-z$ views, respectively.  This format makes it convenient to specify any pixel in terms of layer, face, module, ROC, and pixel row and column on the ROC.  The entire pixel detector is immersed in a uniform solenoidal magnetic field oriented along the $z$-axis, resulting in charged particle trajectories that are curved in the $R-\phi$ (\emph{bend}) view and straight in the $R-z$ (\emph{non-bend}) view.

\begin{figure*}[hbp]
\centering
\includegraphics[width=0.98\textwidth]{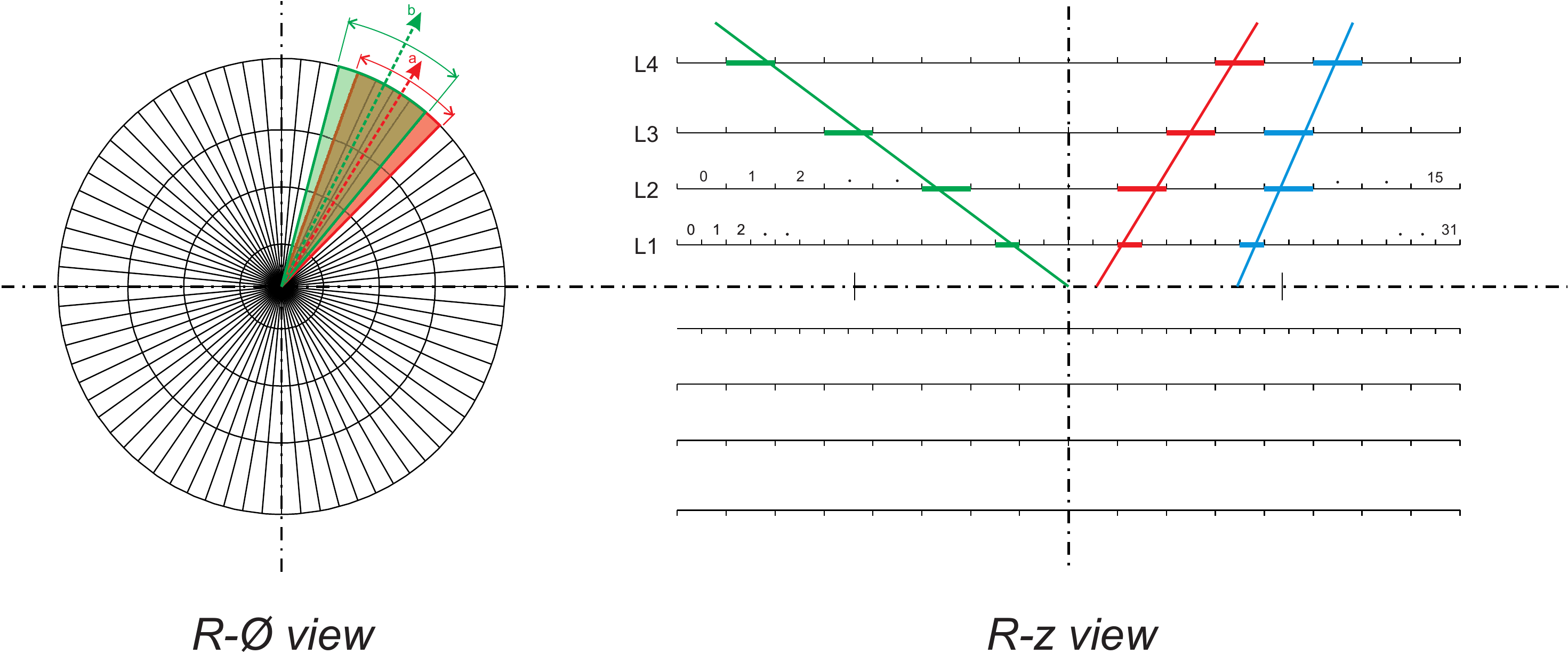}
\caption{The toy detector is divided into 72 overlapping sectors in the $R-\phi$ view.  Examples of two such neighboring sectors, the 12th in red and the 13th in green, each measuring $25^{\circ}$ are shown above.  In the $R-z$ view, each layer is divided into equal sections in $z$ --- 32 for layer 1 and 16 for all other layers.  Shown in blue, red, and green are examples of combinations of 4 sections (1 from each layer) forming roads of straight tracks originating from the luminous region and within the fiducial volume of the detector. }
\label{fig:sector_roads}
\end{figure*}

\begin{figure*}[htbp]
\centering
\subfloat[]{
  \includegraphics[clip,width=0.406\textwidth]{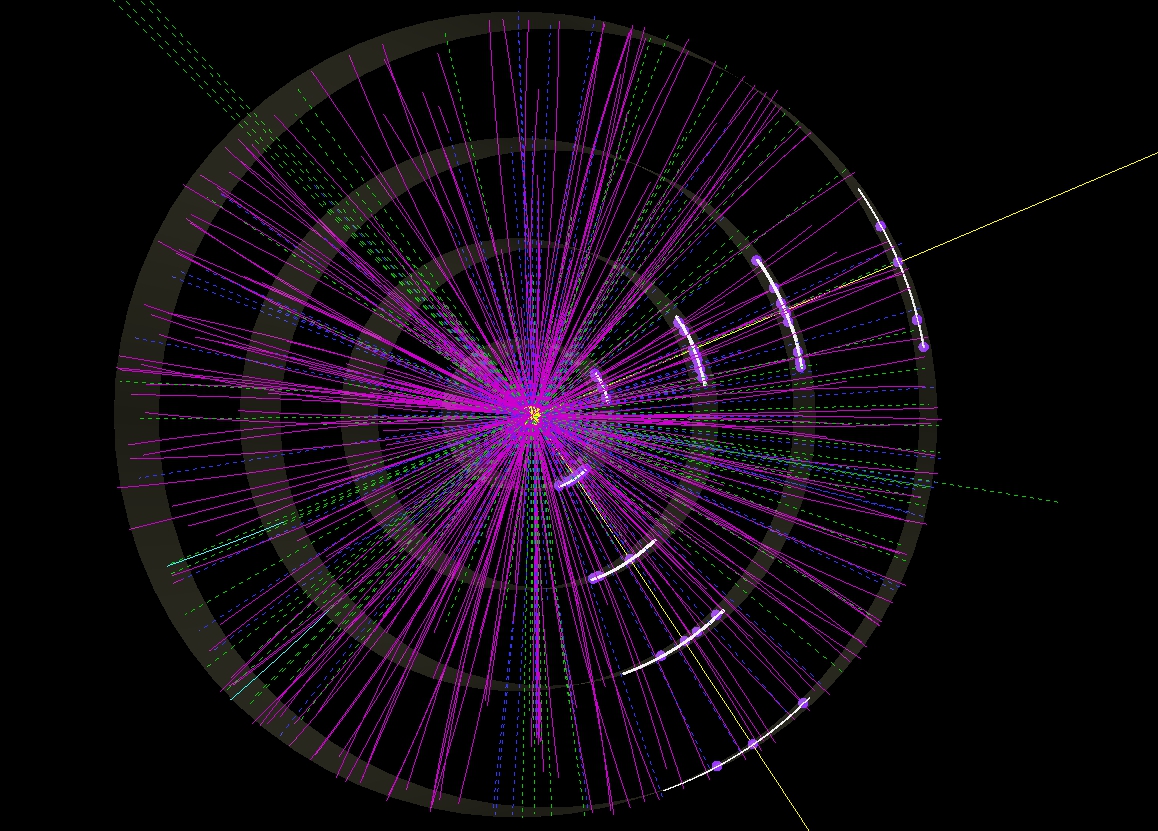}
}
\subfloat[]{
  \includegraphics[clip,width=0.58\textwidth]{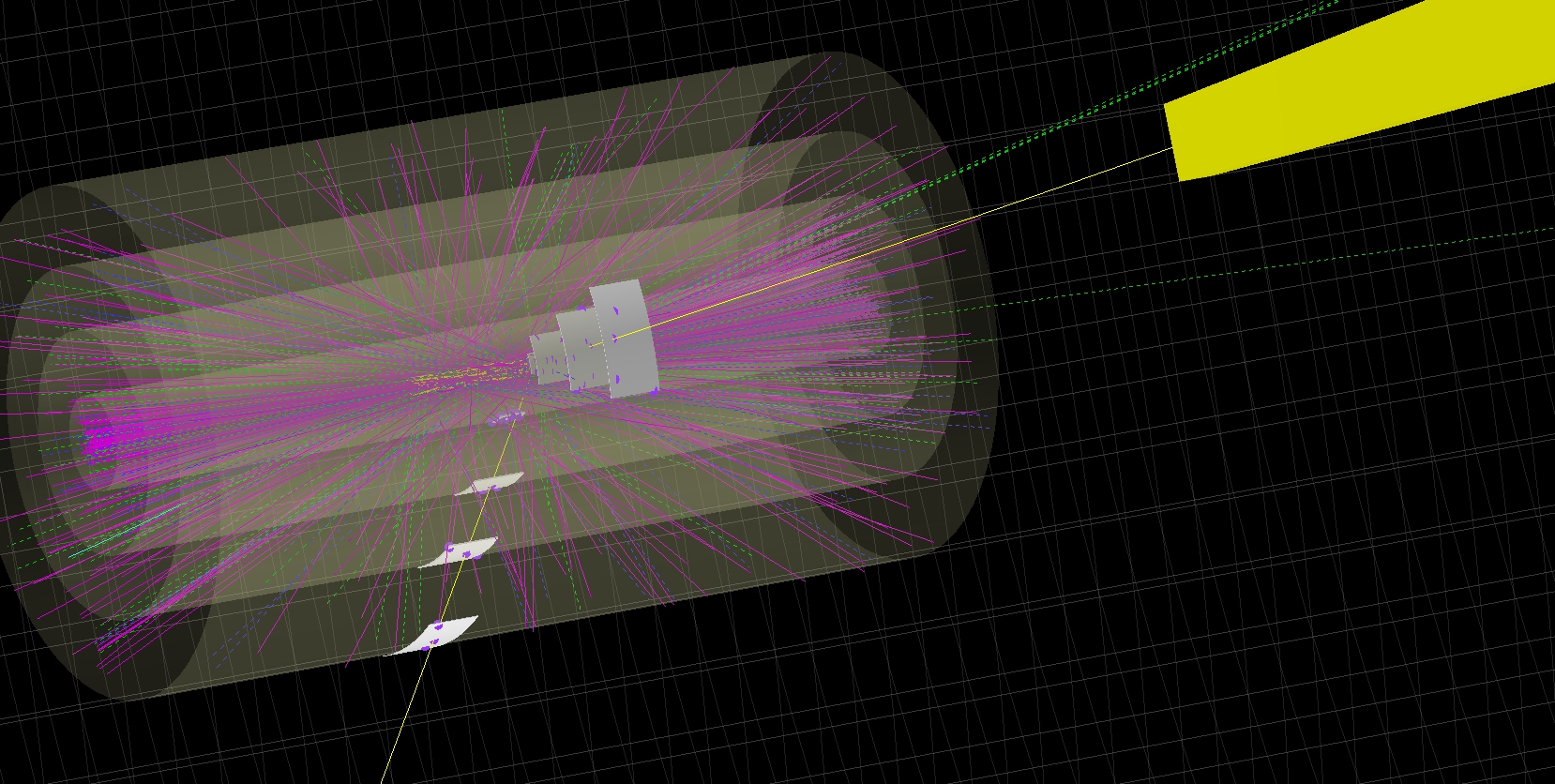}\
}
\caption{ Two views of an event with 2 EM calorimeter clusters (one of which is represented by the yellow bar in (b)) above the threshold.   The search for tracks associated with the clusters is done using only hits in the region-of-interest (ROI) defined by the 4 curved white tiles.  There are 2 ROIs in the figure, each pointing at an EM cluster.}
\label{fig:roitiles}
\end{figure*}

For the purposes of defining the regions of interest (ROI) used in detector data readout and defining the pattern banks used by the automata processor (both discussed in more detail below), we divide the pixel detector into logical sections in the $R-\phi$ and $R-z$ views (see Figure~\ref{fig:sector_roads}).  In the $R-\phi$ view, the detector is divided into 72 overlapping azimuthal sectors, each measuring $25^{\circ}$ in $\phi$.  In the $R-z$ view, layer 1 is divided into 32 equal sections in $z$ and each of layers 2, 3, \& 4 into 16 equal sections in $z$.  Taken together, the logical divisions in both views form cylindrical tube segments in each layer which we will refer to as \emph{tiles}.  An example of these tiles being used to define the ROIs for track searching is shown in Figure~\ref{fig:roitiles}.

Surrounding the pixel detector is a larger concentric cylinder representing the barrel portion of the EM calorimeter with a radius of 129 cm and eta coverage of $\pm1.479$.  We assume a single crystal barrel granularity of 180-fold in $\phi$ (half that of the CMS detector) and $2\times85$ fold in $\eta$.

\begin{figure*}[htbp]
\centering
\subfloat[]{
  \includegraphics[width=0.517\textwidth]{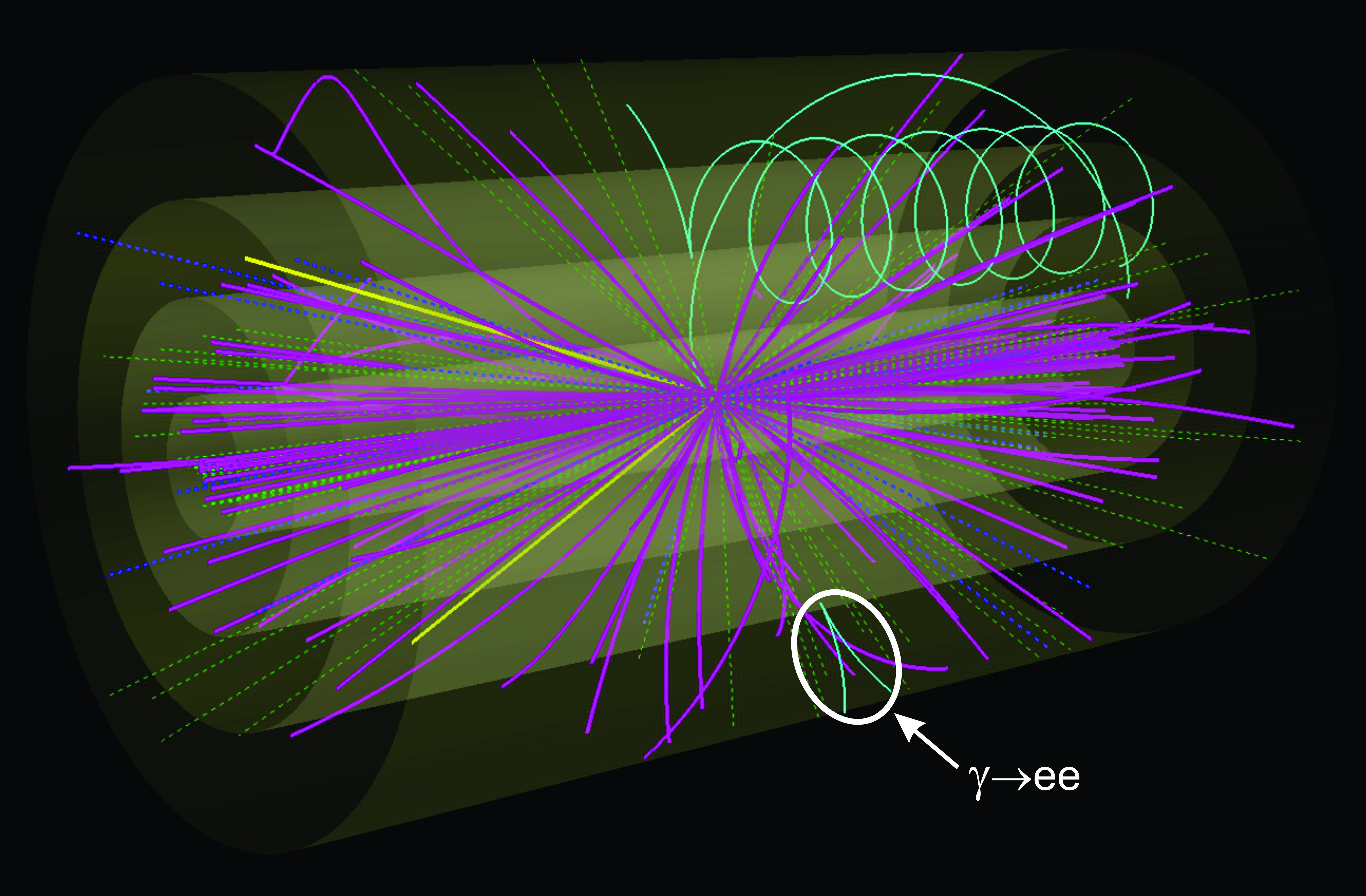} 
}
\subfloat[]{
  \includegraphics[width=0.469\textwidth]{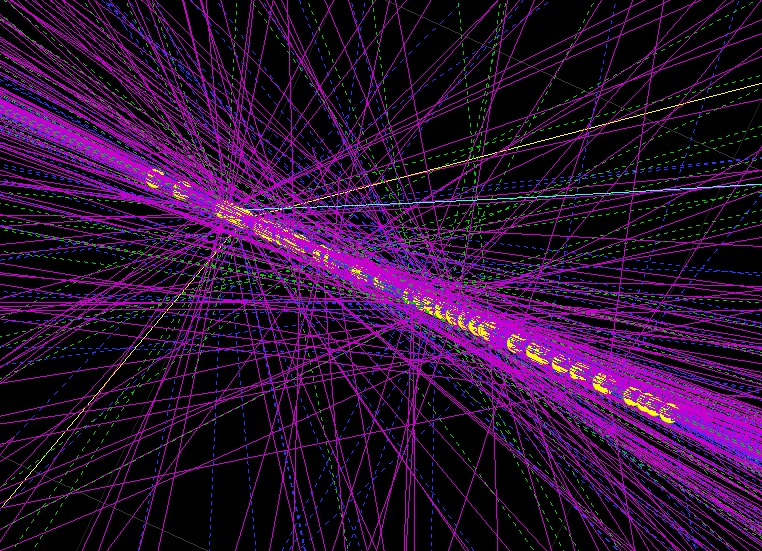} \
}
\caption{A simulated event consisting of a single $Z\rightarrow ee$ interaction is shown in (a).  The two yellow tracks are the electrons from the $Z$.  Other electrons are drawn in cyan.  All other charged tracks are drawn in magenta.  Photons are drawn as green dashed lines and all other neutral tracks are drawn as blue dashed lines.  The downward-pointing narrow opening angle pair drawn in cyan is from a photon conversion in the 4th layer.  The image in (b) shows a simulated event with a $Z\rightarrow ee$ interaction that is overlaid with 50 pileup interactions. }
\label{fig:gensim}
\end{figure*}

\subsection{Simulated Events}\label{sec:gensim}
Simulated events are generated with proton-proton collisions at 14 TeV center-of-mass energies using Pythia 6.4~\cite{ref:pythia64} with CMS Tune Z2*~\cite{ref:tunez} parameters.  Each event consists of a $Z\rightarrow ee$ signal interaction overlaid with pileup interactions which are pure Monte-Carlo minimum bias interactions.  The number of pileup interactions overlaid on each signal interaction is randomly chosen from a Poisson distribution.  The positions of the primary interaction vertices along the $z$-axis are randomly selected from a Gaussian distribution  centered at $z=0$ cm with $\sigma_z=5$~cm.  For simplicity, the transverse positions of the primary vertices are always centered at $x=y=0$~cm. Figure~\ref{fig:gensim}-a shows a 3D view of a simulated event consisting only of a single $Z\rightarrow ee$ signal interaction.  Figure~\ref{fig:gensim}-b shows an event with a $Z\rightarrow ee$ interaction overlaid with 50 pileup interactions.

Starting from the production vertices, all particles are tracked through the toy detector model with a uniform 4T axial magnetic field.  All unstable particles are allowed to decay randomly into a channel selected according to their branching fractions and with exponential decay length distributions based on their proper lifetime.  Photons are converted into electron-positron pairs with the appropriate small opening angle distribution at a frequency determined by the pair production cross section (for both nuclear and electronic fields) in the material of the detector concentrated in the 4 pixel detector layers.  We use a fictitious \emph{sensor module} material with an effective Z and density ($\rho$) determined from the various components of an actual CMS pixel sensor module.  To keep things simple in this proof-of-concept investigation, no other physics processes such as multiple scattering and energy loss mechanisms are simulated.  All charged particles traversing a pixel detector layer generate a single pixel hit for that layer with 100\% efficiency.  The address of the hit is determined from the coordinates of the intersection of the trajectory with the cylinder.   In reality, a charged particle can produce a cluster of neighboring hits in a layer, which could lead to more fake tracks in the pattern recognition stage. This is not included in our simplified simulation. Because our selection criteria requires EM clusters to have $p_T\geq5$ GeV, only electromagnetic particles (electrons and photons) above this threshold are propagated beyond layer 4 of the pixel detector to the EM calorimeter barrel.  Upon reaching the calorimeter, the total energy of such an EM particle is deposited into the calorimeter crystal it intersects to generate an EM cluster.

Four different 1000-event samples were generated having Poisson means for the number of overlaid pileup interactions of $N=50$, $80$, $110$, \& $140$.  All four samples used the same set of  $Z\rightarrow ee$ signal interactions.

\section{Implementing the Track Finder on the Automata Processor}

\subsection{Basic Automata Network and Principle of Operation}

\begin{figure*}[htbp]
\centering
\includegraphics[width=0.98\textwidth]{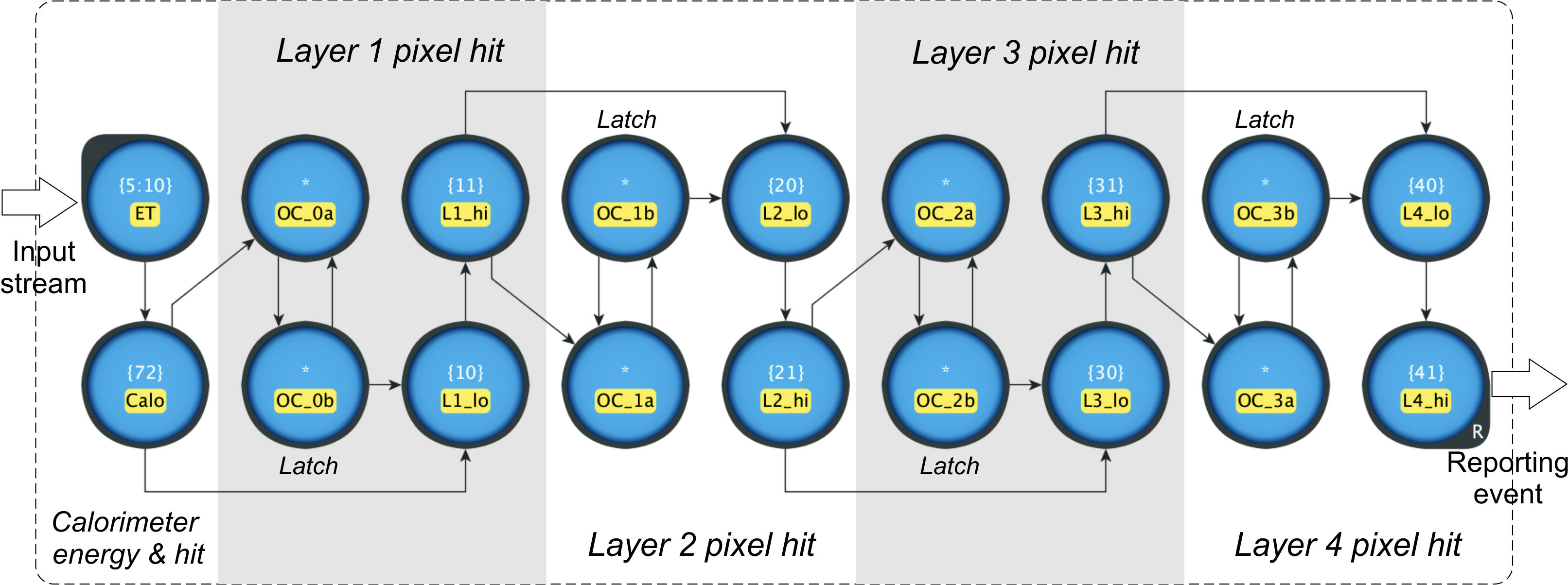}
\caption{An automata network programmed to generate a report on matching a sparse and ordered sequence of exactly 4 pixel hit addresses. The ``R" on the bottom right STE indicates that it is a reporting element.}
\label{fig:all4calo_16bit}
\end{figure*}

The application of the automata processor as a fast pattern search engine is based on the idea of maintaining a database or dictionary of all possible patterns against which symbols in an input data stream are simultaneously compared for matches~\cite{ref:amristori,ref:amchip03}.  The patterns in this dictionary can be all the possible words in a language or, in our case, all physically possible charged particle trajectories in a tracking detector.  In our present case, each possible trajectory is defined by the addresses of 4 pixel hits representing the intersection of the trajectory with each layer of the detector. On the automata processor, each pattern of 4 pixel addresses representing a trajectory in the dictionary is associated with the automata network having the topology shown in Figure~\ref{fig:all4calo_16bit}.  As shown, this network consists of 9 columns of 2 STEs each.  The pixel addresses are stored in the STE pairs labeled $Lx\_lo$ and $Lx\_hi$ with \emph{x=1,2,3,4} denoting the detector layer.  Since each STE only has 8 bits of symbol recognition capability, the least and most significant bytes of each address are stored separately in $Lx\_lo$ and $Lx\_hi$, respectively.  The STE pair to the left of each $(Lx\_lo, Lx\_hi)$ pair acts as a latch that re-enables the $Lx\_lo$ STE on its right only on odd clock cycles.  This ensures that pair of bytes denoting a pixel address are contiguous and can occur anywhere in the input data stream.  The very first pair of STEs on the left of the figure imposes additional constraints for each pattern.  Assuming that the very first symbol in an input data stream represents the energy of the EM cluster for which we are trying to find a track match, the STE labeled $ET$ requires this energy to be within the possible range for the stored track pattern.  Below this, the STE labeled $Calo$ holds the address of the EM calorimeter crystal that trajectory intersects.

To illustrate the operation of this automata network, consider an input data stream from the detector, consisting of pixel hit addresses read out sequentially by layer, starting from the innermost.  Assume that the very first symbol in the input data stream is an 8-bit number representing the EM calorimeter cluster energy and that the second symbol represents the address of the EM calorimeter crystal associated with the cluster. If the energy of the measured cluster falls within the range of the $ET$ STE, its output is enabled, thereby activating the $Calo$ STE.  If the crystal address matches that in the $Calo$ STE, this STE enables its output and activates the \emph{L1\_lo} STE.  After this point, if a symbol anywhere in the data stream matches that in \emph{L1\_lo}, the output of this STE is enabled, activating \emph{L1\_hi}.  If the immediately following symbol matches that in \emph{L1\_hi}, then \emph{L2\_lo} is activated to wait for matching hits in the 2nd layer of the detector.  On the other hand, if no match is found for \emph{L1\_hi}, we need to keep reactivating \emph{L1\_lo} every other clock cycle until we find a layer 1 match.  This is the purpose of the \emph{OC\_0x} pair of STEs which reactivates \emph{L1\_lo} on every odd clock cycle, re-initializing the search for the byte-pair representing the layer 1 hit.  These steps are repeated for every detector layer until either 4 matching pixel hits are found or the input data stream is exhausted.

\subsection{Generating the Dictionary or Pattern Banks}\label{sec:patbank}

\begin{figure}[htbp]
\centering
\includegraphics[width=0.48\textwidth]{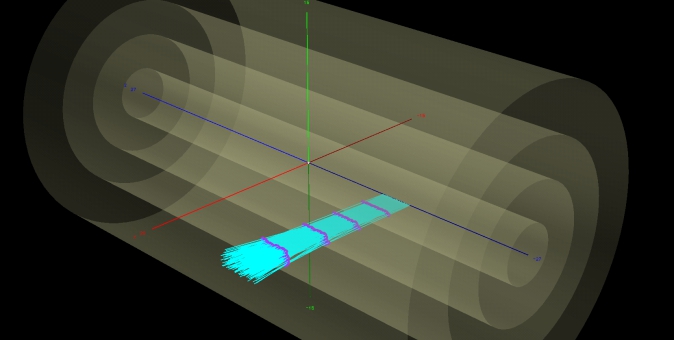}
\caption{Generating all possible track patterns with a single electron gun monte carlo event generator for a given region of the detector.}
\label{fig:patternbanktest}
\end{figure}

We generate our pattern banks separately for the $R-\phi$ bend view and for the $R-z$ non-bend view~\cite{ref:3dpats}.  For the $R-\phi$ view, the patterns are generated for each $\pm12.5$ degree $\phi$-sector described earlier in Section~\ref{sec:bpixtoy}.  For the $R-z$ view, we start with the same subdivisions along $z$ in each detector layer described in Section~\ref{sec:bpixtoy}.  We will refer to each division as a \emph{window}.  We take all possible combinations of 4 windows (one from each layer) forming roads containing straight lines originating from the luminous region and ending in the EM calorimeter barrel.  Patterns are then generated for each road.

\begin{figure*}[htbp]
\centering
\includegraphics[width=0.98\textwidth]{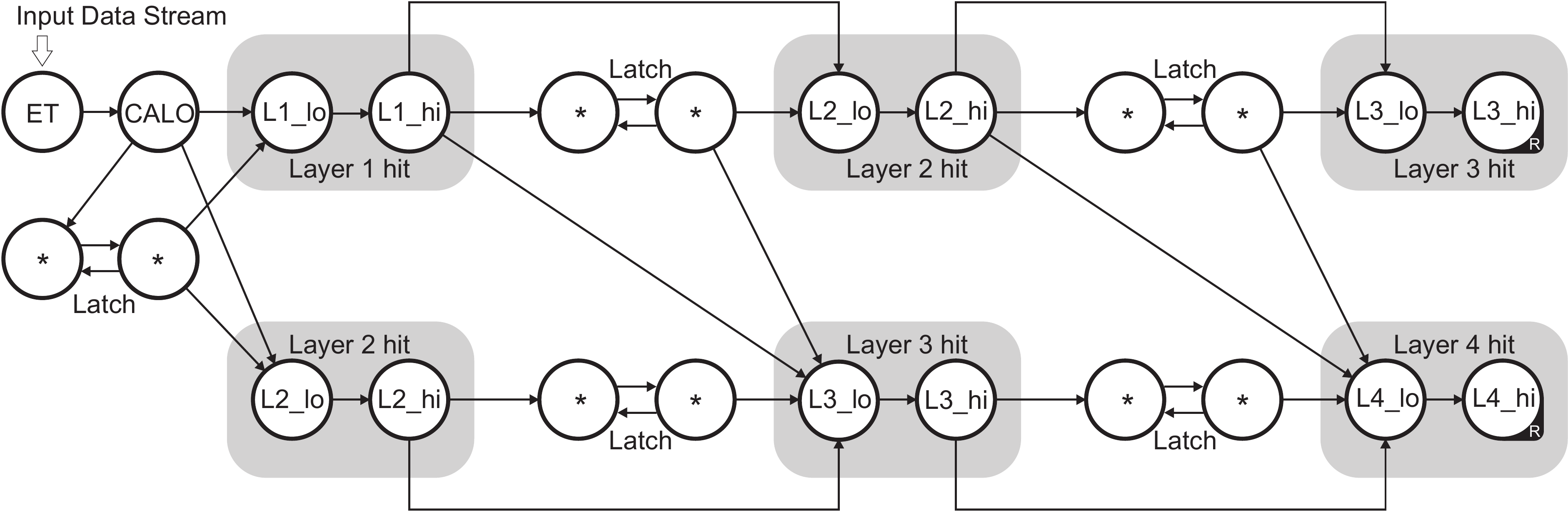}
\caption{An automata network programmed to generate a report on matching a sequence of 4 pixel hit addresses allowing up to one missing hit. The ``R"s on the two rightmost STEs indicate that these are reporting elements.}
\label{fig:majority_16bit}
\end{figure*}

To generate the pattern banks, we use a single electron gun monte carlo event generator to produce all the possible tracks above $p_T=5$ GeV.  Figure~\ref{fig:patternbanktest} shows the process of generating these patterns for a given region of the detector. Electrons of both charges are generated and propagated through the toy detector model described in Section~\ref{sec:bpixtoy} generating pixel hits as they traverse each layer.  Just as with our simulated $Z\rightarrow ee$ samples that are overlaid with  pileup interactions, we do not simulate energy loss or multiple scattering when generating the pattern banks.

To keep the total number of patterns manageable, we use a resolution of 4 pixels for all detector layers in each view. There is a total of 72 pattern banks in the $R-\phi$ view with each bank having $\sim$1163 unique track patterns. Due to the rotational symmetry of the ideal toy detector about the $z$ axis, each bank has an identical set of patterns modulo a rotation in $\phi$. One could, in principle, use a single bank to represent the entire detector in the $R-\phi$ view after the appropriate rotation is applied to the hits of a given sector.  We chose not to do this since perfect rotational symmetry may not be present in a real detector. For the $R-z$ view, we consolidate all roads with a common layer 1 and layer 4 window into a single bank resulting in a total of 244 pattern banks.  The average number of track patterns in each bank is $\sim$4662.  For the ideal toy detector, there is a mirror symmetry about the $x-y$ plane in this view.  The set of track patterns in each half are identical modulo a translation in $z$.  Again, we chose not to exploit this symmetry since it may not be present in a real detector.

These pattern banks are programmed into the automata processor using the C-API provided in the AP SDK.  For each entry in a generated pattern bank, an instance of a macro representing the basic automata network shown in Figure~\ref{fig:all4calo_16bit} is created by substituting its parameters with the values associated with the current pattern. The resulting automata network containing all instances is then compiled into an object file that is loaded into the automata processor.

We can compile 2,496 instances of the macro shown in Figure~\ref{fig:all4calo_16bit} onto the current version of the AP chip.  Without taking advantage of rotational or translational symmetries, this would require about 34 chips for the $R-\phi$ view and 456 chips for the $R-Z$ view.  With the current AP boards that have 4 ranks of 8 chips each, this translates to about 1 board for the $R-\phi$ view and 14 boards for the $R-Z$ view.

\subsection{Available Features and Capabilities and Possible Improvement}

The simple automata network shown in Figure~\ref{fig:all4calo_16bit}, requiring hits in all 4 layers of the detector, suffices for the present purpose of demonstrating a proof of principle.  However, it is possible to design an automata-based algorithm that can deal with missing hits due to inefficiencies in a real detector.  An automata network representing a track pattern allowing up to one missing hit in any layer is shown in Figure~\ref{fig:majority_16bit}.  In the general case when more than 1 missing hit is allowed, the total number of STEs representing a pattern is given by $N_{ste}=2(2N_{l}+2N_{l}N_{m}-3N_{m}-1)$ where $N_l$ and $N_m$ are the number of detector layers and number of allowed missing hits, respectively.  When there are no missing hits, the total number of STEs is simply $N_{ste}=4N_{l}+2$.

Another interesting feature of the AP is the STE's ability to recognize a range of values instead of a specific one.  This makes it possible to employ variable resolution patterns that offer a way to reduce the total number of patterns in a bank~\cite{ref:vrescam}.

Lastly, the number of STE's (18) in Figure~\ref{fig:all4calo_16bit} needed to represent a specific track pattern is largely due to the limited alphabet size of 8 bits.  If our alphabet size were 16 bits, we could reduce the number of STEs by a factor of 3 to just 4 STEs representing the 4 hits plus two additional STEs for the calorimeter energy and position.  This would reduce the total number of automata chips needed for our pattern banks.

\begin{table*}[t]
\centering
\begin{center}
\begin{tabular}{@{\extracolsep{4pt}}S[table-format=3]S[table-format=4]S[table-format=3]S[table-format=3]S[table-format=3]S[table-format=2]S[table-format=3]S[table-format=2]S[table-format=2] }
\hline \hline
{Pileup} & \multicolumn{3}{c}{EM Clusters} & \multicolumn{2}{c}{Track Match} & {Eff.} & {Rejection} &{Purity} \\
\cline{2-4}\cline{5-6}
{Inter.} & {Total} & {$e$}  & {$\gamma$} & {$e$} & {$\gamma$}  & {(\%)} & {Factor} & {(\%)}\\
\hline
 50   & 1242 & 837 & 405 & 837 &   9 & 100 & 45 & 99 \\ 
 80   & 1395 & 839 & 556 & 839 & 17 & 100 & 33 & 98\\
110  & 1515 & 844 & 671 & 844 & 26 & 100 & 26 & 97\\
140  & 1648 & 844 & 804 & 844 & 56 & 100 & 14 & 94\\
\hline\hline
\end{tabular}
\end{center}
\caption{This table summarizes the ability of the Automata Processor algorithm to identify electrons and reject photons for each of the 4 simulated samples used in our study.  The total number of EM clusters and their breakdown into electrons and photons are shown in columns 2-4.  Columns 5 and 6 show the number of electron clusters and photon clusters, respectively, for which a matching track was found.  The last three columns show the electron identification efficiency, photon rejection factor, and purity as defined in the text.}
\label{tab:trigeff}
\end{table*}

\section{Testing the AP-based Electron Confirmation Trigger with the Simulated Samples}
To satisfy the requirements of the electron track trigger application described in Section~\ref{sec:poc}, the AP must, first of all, be able to reconstruct tracks for electron/photon discrimination.  Secondly, it must be able to accomplish this task within the available latency of the trigger~\cite{ref:latency}. This section will focus on its track finding ability.  The next section will focus on its processing times.

\subsection{Testing Procedure}
For each of the simulated events described in Section~\ref{sec:gensim}, we check to see if there are EM calorimeter clusters.  Selecting only EM clusters with $p_T>5$ GeV, we then read out all the pixel hits within the ROI associated with the cluster (see Figure~\ref{fig:roitiles}).  The extents of the ROI in the $R-\phi$ view are defined by the boundaries of the $\phi$ sector (1 of 72 described in Section~\ref{sec:bpixtoy}) whose bisector is closest in azimuthal distance to the cluster.  Considering only hits within the ROI is a sensible and practical approach in an actual implementation because it avoids having to deal with the sheer amount of data involved in reading out the entire pixel detector.

In this proof-of-principle study, we also assume single (double) crystal calorimeter resolution in the $R-Z$ ($R-\phi$) view and precise knowledge of the interaction vertex associated with the $Z\rightarrow ee$ signal.  For the purposes of this study, we assume that the latter knowledge is provided by an independent sub-detector system such as an outer tracker based on silicon strips. In the $R-z$ view, we first construct a straight line defined by the calorimeter cluster coordinates (center of the single crystal) and the primary interaction vertex.  We then find the layer 1 and layer 4 \emph{windows} this line intersects.  Together with this pair, all layer 2 and layer 3 windows that form roads (see Section~\ref{sec:patbank}) with the pair are used to define the ROI extents in the $R-z$ view.  If such a ROI exists within the acceptance of the pixel detector, we will refer to the EM cluster plus primary vertex pair as \emph{reconstructable}. All the pixel hits within the ROI defined this way are read out sequentially by layer starting with the innermost.  The sequence of the hits within each layer in the data stream does not matter to the automata track finding algorithm.  This data stream of pixel hits arranged by layer is appended to two 8-bit quantities, the first for the EM cluster energy and the second for the EM calorimeter crystal coordinate.  The 8-bit stream is fed into the automata hardware containing the pattern banks corresponding to the ROI.  This is done separately for the $R-\phi$ and $R-z$ views.  In the hardware, all the instances of the macro represented by Figure~\ref{fig:all4calo_16bit} associated with the bank are simultaneously presented with the input data stream to generate reports in case of matches.  A trigger accept is generated if matches are reported in both views on the same clock cycle.

\subsection{Results for Electron Identification and Photon Rejection}

All reconstructable EM clusters are required to have $p_T>5$ GeV and to originate from the beam axis. The total number of EM clusters satisfying these criteria for each of the 1000-event simulated samples overlaid with a different number of pileup interactions are shown in the second column of  Table~\ref{tab:trigeff}. A breakdown of these numbers into those originating from electrons and those from photons is shown in the third and fourth columns.  The fifth column of the table shows the number of electron clusters for which a matching track was found.  The same number for photons is shown in the sixth column.

For this study, we define the electron identification efficiency as $\epsilon_{\mathrm{e}}=N_{\mathrm{EMtrk}}^\mathrm{e}/N_{\mathrm{EM}}^\mathrm{e}$ and the photon rejection factor as 
$R_{\mathrm{\gamma}}=N_{\mathrm{EM}}^\mathrm{\gamma}/N_{\mathrm{EMtrk}}^\mathrm{\gamma}$. $N_{\mathrm{EM}}^\mathrm{e}$ and $N_{\mathrm{EM}}^\mathrm{\gamma}$ are the number of reconstructable EM clusters originating from electrons and photons (columns 3 and 4 of Table~\ref{tab:trigeff}), respectively, which satisfy the requirements described at the beginning of this section. $N_{\mathrm{EMtrk}}^\mathrm{e}$ and $N_{\mathrm{EMtrk}}^\mathrm{\gamma}$ are the corresponding numbers of EM clusters for which there is a matching track in the pixel detector (columns 5 and 6 of Table~\ref{tab:trigeff}). The results for $\epsilon_{\mathrm{e}}$ and $R_{\mathrm{\gamma}}$ are shown in the last two columns of Table~\ref{tab:trigeff}. For all samples, we see that the AP algorithm correctly finds a matching track for every electron EM cluster satisfying our requirements.  The fraction of photon EM clusters satisfying our requirements that are misidentified increases from 2\% to 7\% as the number of the pileup interactions in the sample, and hence detector occupancy, increases. The last column in Table~\ref{tab:trigeff} shows the purity $P_{e}=N_{\mathrm{EMtrk}}^\mathrm{e}/(N_{\mathrm{EMtrk}}^\mathrm{e}+N_{\mathrm{EMtrk}}^\mathrm{\gamma})$, which we define as the fraction of all EM clusters satisfying our track trigger requirements that originated from an electron.

\begin{figure*}[htbp]
\centering
\includegraphics[width=0.98\textwidth]{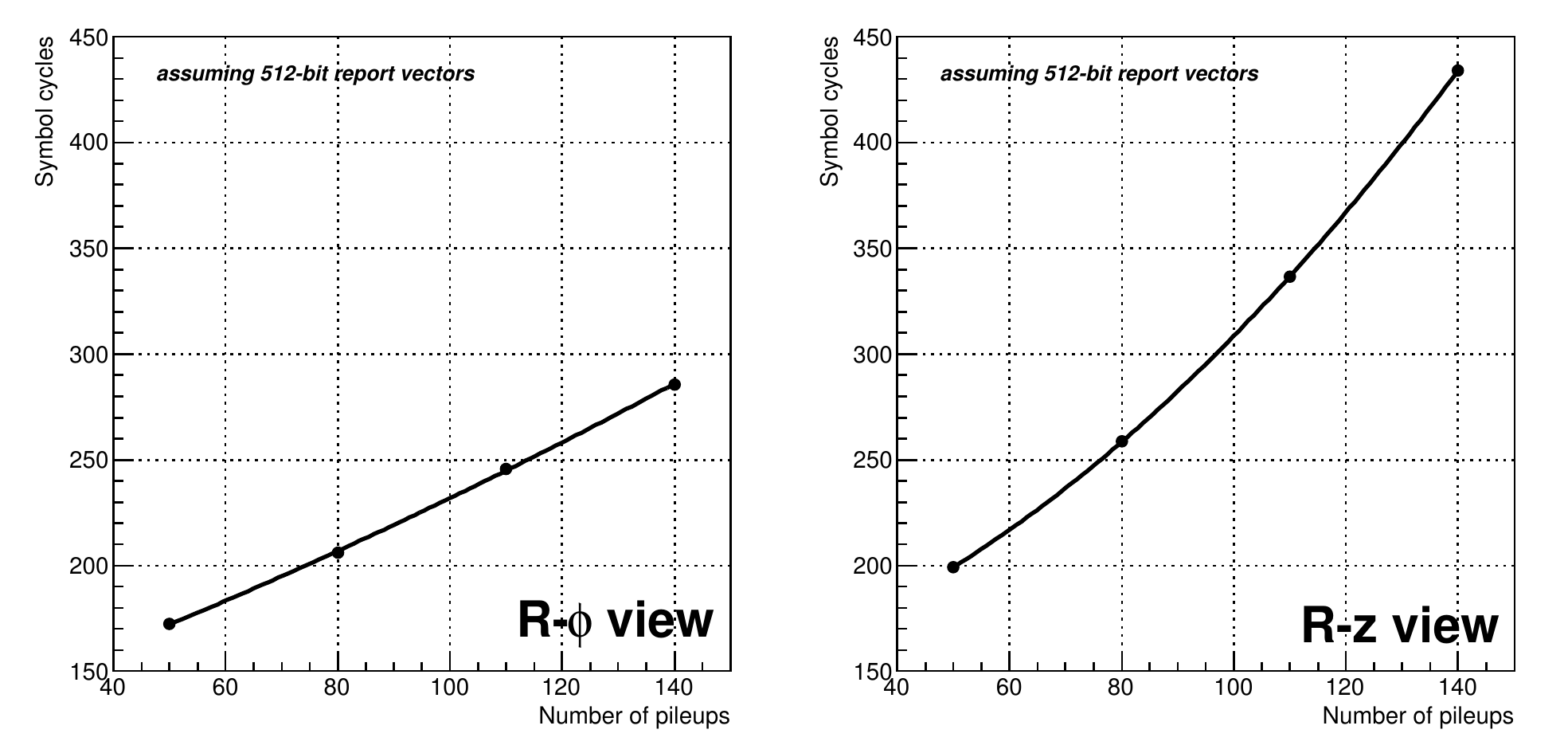}
\caption{Shown above are the symbol cycles to process the hits in each view and transfer the report events to the output event buffers on an AP chip.  This is plotted as a function of the simulated samples overlaid with $N=50$, $80$, $110$, and $140$ pileup interactions.}
\label{fig:apcycles}
\end{figure*}

\section{Processing Time}\label{sec:aptimes}
The amount of time it takes the AP to find matching tracks in each view (which we will refer to as \emph{symbol processing time}) depends only on the number of hits in the ROI.  Because we are using 16-bit hits while the AP uses native 8-bit symbols, the total number of input symbols is twice the number of hits.  Since one symbol is processed per AP clock cycle, the symbol processing time is simply equal to the number of input symbols.  This time does not represent the total processing time since the match results in each view need to be read out of the AP chips and undergo further processing to find coincident matches in each view before making a trigger decision.  This additional time includes (a) the \emph{internal data-transfer time} within the AP to move the match results from the local event memories in the six output regions to the output event buffers (see Section~\ref{sec:apreparch}), and (b) the \emph{external processing time} in external logic to find coincident reports in both views. The internal transfer time depends on the size of the report vectors and their number in each of the six regions.  In addition, there are overheads associated with determining a region to be empty and a startup overhead for reading the first vector. What we refer to as the external processing time includes the data-transfer time from the AP's event output buffer to the external logic across the DDR3 interface. The symbol processing and internal data transfer will be collectively referred to as \emph{core processing} since they both occur on the AP chip.  The first subsection below will focus on the core processing followed by a second subsection devoted to the external processing.

\subsection{Core Processing Time}
To calculate the number of symbol processing cycles, we simply multiply the number of hits by two. For both views, we add an additional symbol cycle to process the 8-bit symbol representing the coordinates of the calorimeter crystal.  For the $R-\phi$ view, we add one more symbol cycle to account for the 8-bit symbol representing the energy of the cluster. 

As we explained earlier, two report events occurring on the same symbol cycle are saved in either the same or different event vectors depending on the output region they occurred in.  As we also pointed out, it takes a finite number of symbol cycles to transfer each vector.  Because of this, when calculating the internal data-transfer time, we randomly assign a report event to one of the six output regions.  This assumes there is no correlation between the automata instances representing all the patterns in our bank and that they are uniformly distributed throughout the 6 regions.  We use an event vector divisor of $2$ to reduce the vector width to 512 bits and assign the appropriate number of symbol cycles to transfer a vector of this size. We also take into account the overheads associated with ``reading" an empty region and transferring the first vector. The symbol processing times plus internal data-transfer times for each view are shown as a function of the simulated sample in Figure~\ref{fig:apcycles}.  For a symbol cycle of $7.5$ ns, these translate to $1.29$, $1.55$, $1.85$, and $2.15$ $\mu$s, respectively, for the 50, 80, 110, and 140 pileup samples in the $R-\phi$ view. The corresponding times for the $R-Z$ view are, respectively, $1.49$, $1.94$, $2.53$, and $3.25$ $\mu$s. 

\subsection{External Processing Time}
After all detector hit data associated with an EM cluster are processed by the AP banks for the $R-\phi$ and $R-Z$ views, the last step of our track finding algorithm requires checking for at least one pair of reporting events (one from each view) in which the $R-Z$ event occurred one symbol cycle ahead of the $R-\phi$ event.  This 1-cycle difference is due to the fact that the automaton representing a track in the $R-Z$ view has one less STE than that for the $R-\phi$ view.   This requirement does not guarantee a common source for both events, but those that do originate from the same track will necessarily exhibit this correlation in time, and this constraint can only help reduce fake rates.

\begin{figure*}[htbp]
\centering
\subfloat[]{
  \includegraphics[width=0.45\textwidth]{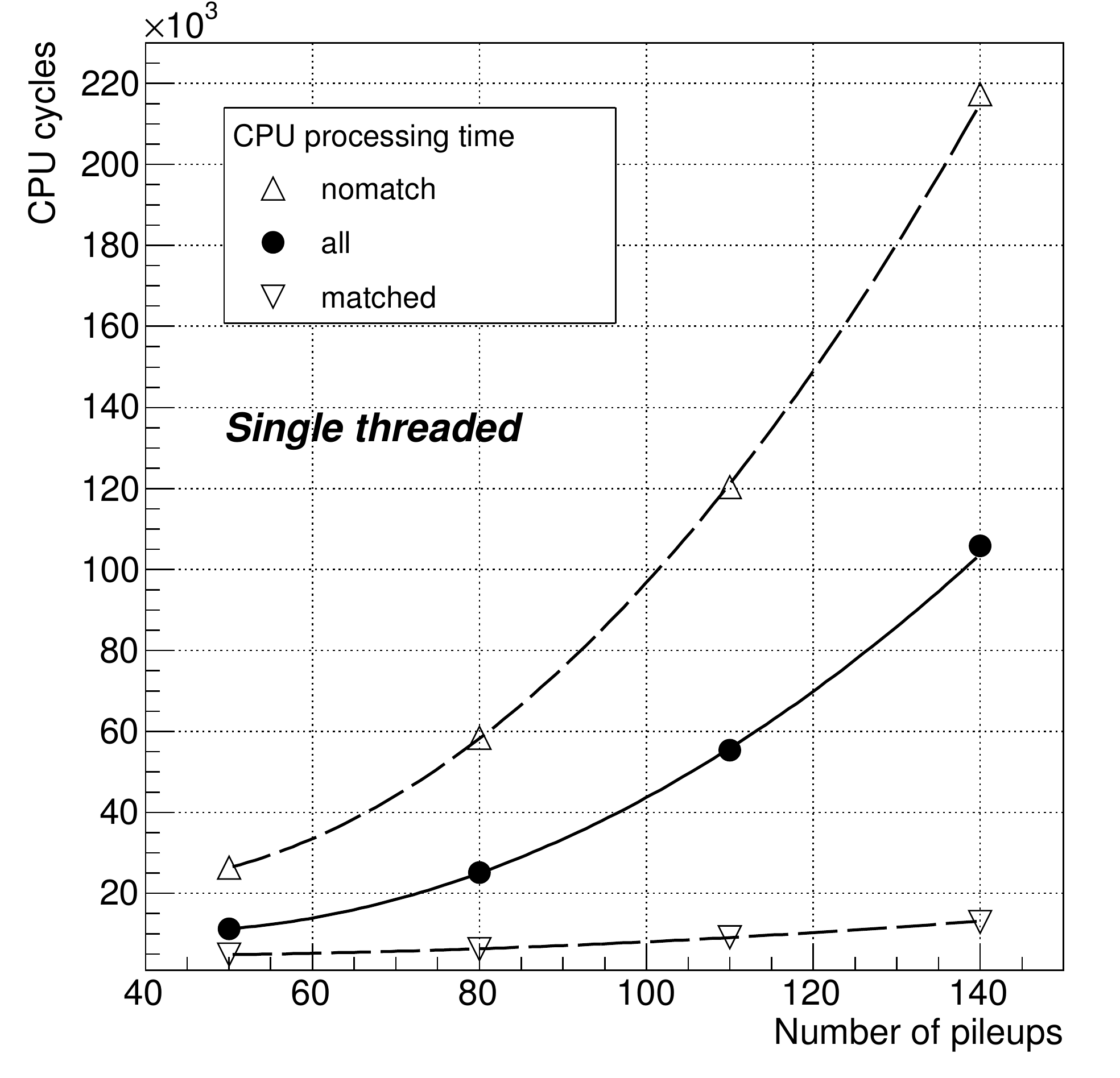}
}
\subfloat[]{
  \includegraphics[width=0.45\textwidth]{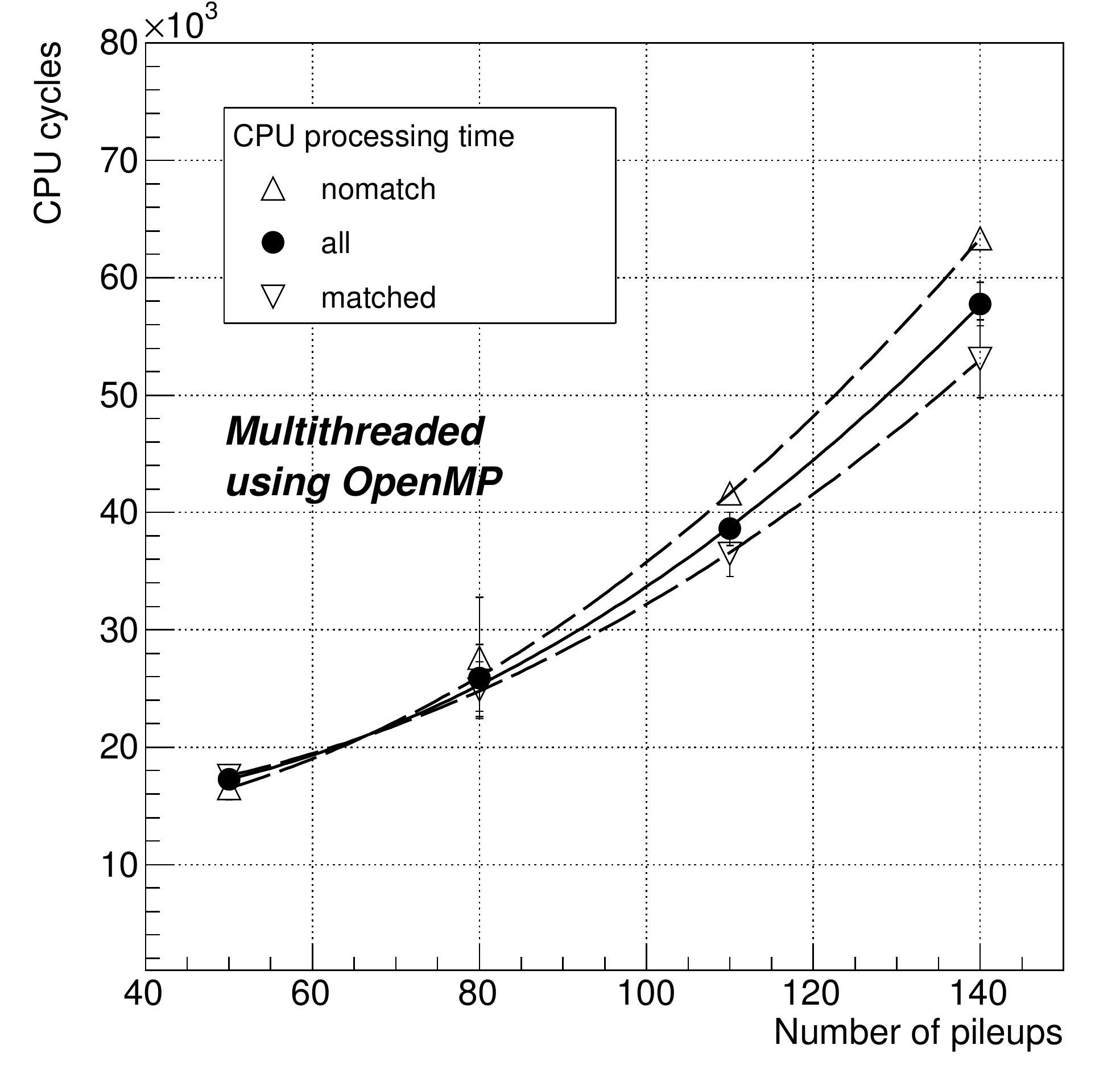}
}
\caption{The plots above show the number of cycles it takes an Intel Core i7 CPU to process the pixel track trigger algorithm described in Ref.~\cite{ref:pixtrk}.  Plots (a) \& (b) show the number of CPU cycles for the single threaded and multithreaded cases, respectively, as a function of the simulated data samples. Dashed-line plots with upright triangular markers are obtained using only EM clusters for which no track match is found. Dashed-line plots with inverted triangular markers are obtained using only EM clusters that have at least one matching track. The solid-line plots and solid circular markers are obtained using all EM clusters.  The cycles in these plots are measured using the Intel CPU's time stamp counter.}
\label{fig:x86cycles}
\end{figure*}

To estimate the processing time associated with finding at least one pair of correlated reporting events, we implemented the Content Addressable Memory (CAM) reference design described in Reference~\cite{ref:xilinx} on a 100 MHz Xilinx Virtex-6 LX240T Field Programmable Gate Array (FPGA).  Our studies are based on simulations done using the Xilinx ISE Design Suite.  We assume that the report vectors can be read out of the Micron AP's 8-bit wide bus at 1066 MHz. In our simulations, the 576-bit data (64-bit header + 512-bit reduced-size vector) associated with each vector is read out from the $R-Z$ banks and the 64-bit header containing the temporal information is extracted and written into FPGA memory.  Once the headers from all $R-Z$ view vectors associated with an EM cluster are stored in memory, we loop over the corresponding headers from the $R-\phi$ view vectors and present them to the CAM to find the first match.  As soon as this is found, a trigger accept signal is generated and the process is repeated for the next EM cluster.

For the 1000-event sample overlaid with 140 pileup interactions, the average time to execute this match finding step on the FPGA is 0.37 $\mu$s, which is insignificant compared to the core processing time.

\section{Comparison with other Processor Architectures}

In order to put the AP in better perspective, we compare it with other processor architectures.  We implement the pixel-based tracking trigger algorithm described in Ref.~\cite{ref:pixtrk} on a CPU and a GPU.  For this algorithm, which is functionally equivalent to that used on the AP, we assume the same detector layout and parameters, apply the same energy threshold cuts, and look at an identical set of hits contained within a ROI defined in the same way.

The CPU and GPU results in Sections~\ref{sec:cputimes} and \ref{sec:gputimes} are presented within the context of the electron confirmation trigger chosen to demonstrate our proof of concept.  The relevant quantity in this case is the total amount of time it takes to process a single event (processing latency) and generate a trigger decision.  In order to avoid dropping events, this must be less than the time available to temporarily store an event prior to a trigger decision, which is dictated by the limited size of the event buffers.  For the CMS experiment, this available time is on the order of $10$ $\mu$s. On the other hand, processing latencies are not as important for other applications like offline reconstruction or online triggers with sufficiently large event buffers. Although such applications may still demand high processing rates, they can easily be satisfied by the addition of more parallel execution units. In Section~\ref{sec:cpurate} we discuss the CPU performance with this latter case in mind by considering how many processing units (cores) can be used in parallel to achieve the same processing rate as the AP-based system.
We do not do this for the GPU since it is much more difficult to determine and control the mapping of threads (and their organization into blocks) to the 15 streaming multiprocessors of 192 cores each, in order to define some processing unit that can be reliably scaled.

Similar approaches to track finding, based on matching patterns stored in a bank, have been implemented using CAMs or Associative Memories (AMs) on FPGAs and custom Application Specific Integrated Circuits (ASICs)~\cite{ref:amristori,ref:amchip03}.  Since comparisons with such solutions provide a more level playing field than with CPUs and GPUs, we conclude this section by briefly describing a recent implementation of a an CAM-based track finder on an FPGA and compare its capabilities with those of the AP.

\subsection{CPU Comparison}\label{sec:cputimes}

\begin{figure}[htbp]
\centering
\includegraphics[width=0.45\textwidth]{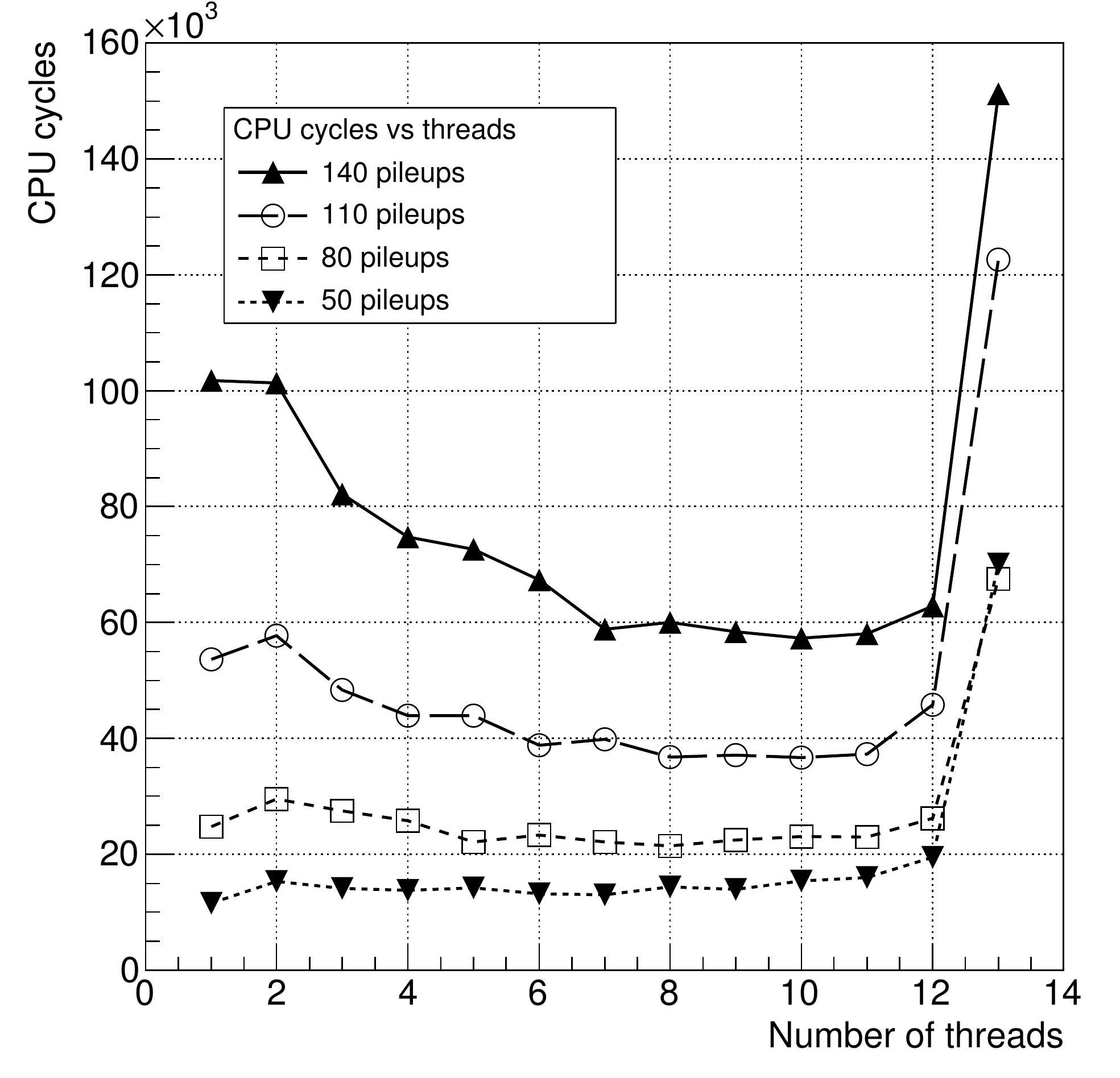}
\caption{The plot above shows the CPU cycles presented in Figure~\ref{fig:x86cycles} as a function of the number of threads for each data sample.}
\label{fig:ompthreads}
\end{figure}

We compile a single threaded C-version of the algorithm described in Ref.~\cite{ref:pixtrk}  using the Intel C compiler (v16.0.1).  We then run it on a 3.3 GHz Intel Core i7 (5820K) processor using the same four simulated data samples.  Using the Intel CPU's time stamp counters, we measure the number of CPU cycles to execute the trigger for each EM cluster and plot the mean as a function of the sample in Figure~\ref{fig:x86cycles}a with a solid line and solid circular markers.  For a $0.3$ ns CPU clock cycle, these results translate to $3.38$, $7.60$, $16.7$, and $32.1$ $\mu$s, respectively, for the 50, 80, 110, and 140 pileup samples.

The processing cycles are also shown separately for EM clusters with at least one matching track (dashed line with upright triangular markers) and EM clusters with no matching tracks (dashed line with inverted triangular markers). Clusters with matching tracks require less processing time because the algorithm quits forming all possible combinations of 4 hits as soon as a matching track is found.  The opposite is true for clusters with no matching tracks since the algorithm ends up attempting all possible combinations.

These results clearly exhibit a quadratic rise in processing times which increases by at least an order of magnitude in going from 50 to 140 pileup interactions. In contrast, the results in Figure~\ref{fig:apcycles} show that the AP processing times scale almost linearly as the number of pileup interactions in the samples is increased.  The processing times increase only by less than a factor of $2\times$ in going from 50 to 140 pileup interactions.  

Using OpenMP, we created a multithreaded version of the code described above and compiled it with the same version of the Intel C compiler.  The results for this multithreaded implementation on the Intel CPU described above, with 6 physical cores (12 logical cores because of Hyper-Threading), are shown in Figure~\ref{fig:x86cycles}b for the four simulated samples. This implementation runs the algorithm using 2 OpenMP threads on each of the 6 physical cores. The meanings of the markers and line types used in the plot are identical to those in Figure~\ref{fig:x86cycles}a.  The average processing times are $5.23$, $7.85$, $11.7$, and $17.5$ $\mu$s, respectively, for the 50, 80, 110, and 140 pileup samples.  The result for the 140 pileup sample is $\sim 2\times$ faster than the single threaded CPU result and $\sim 5\times$ slower than the AP result.
 
The plots in Figure~\ref{fig:ompthreads} show the processing cycles as a function of the number of OpenMP threads for each simulated sample.  These results show that using more threads becomes more advantageous as the number of hits (which increases with number of pileup interactions in the sample) that can be processed in parallel increases. However, as the plot for the 140 pileup sample indicates, using more cores only helps up to a certain point.  Processing performance flattens out beyond 3 physical cores (6 threads) and begins to worsen when we exceed 2 threads per physical core (beyond 12 threads).  Increasing the number of cores beyond a certain point has no effect in reducing the single-event processing time.

\subsection{GPU Comparison}\label{sec:gputimes}

\begin{figure}[htbp]
\centering
\includegraphics[width=0.45\textwidth]{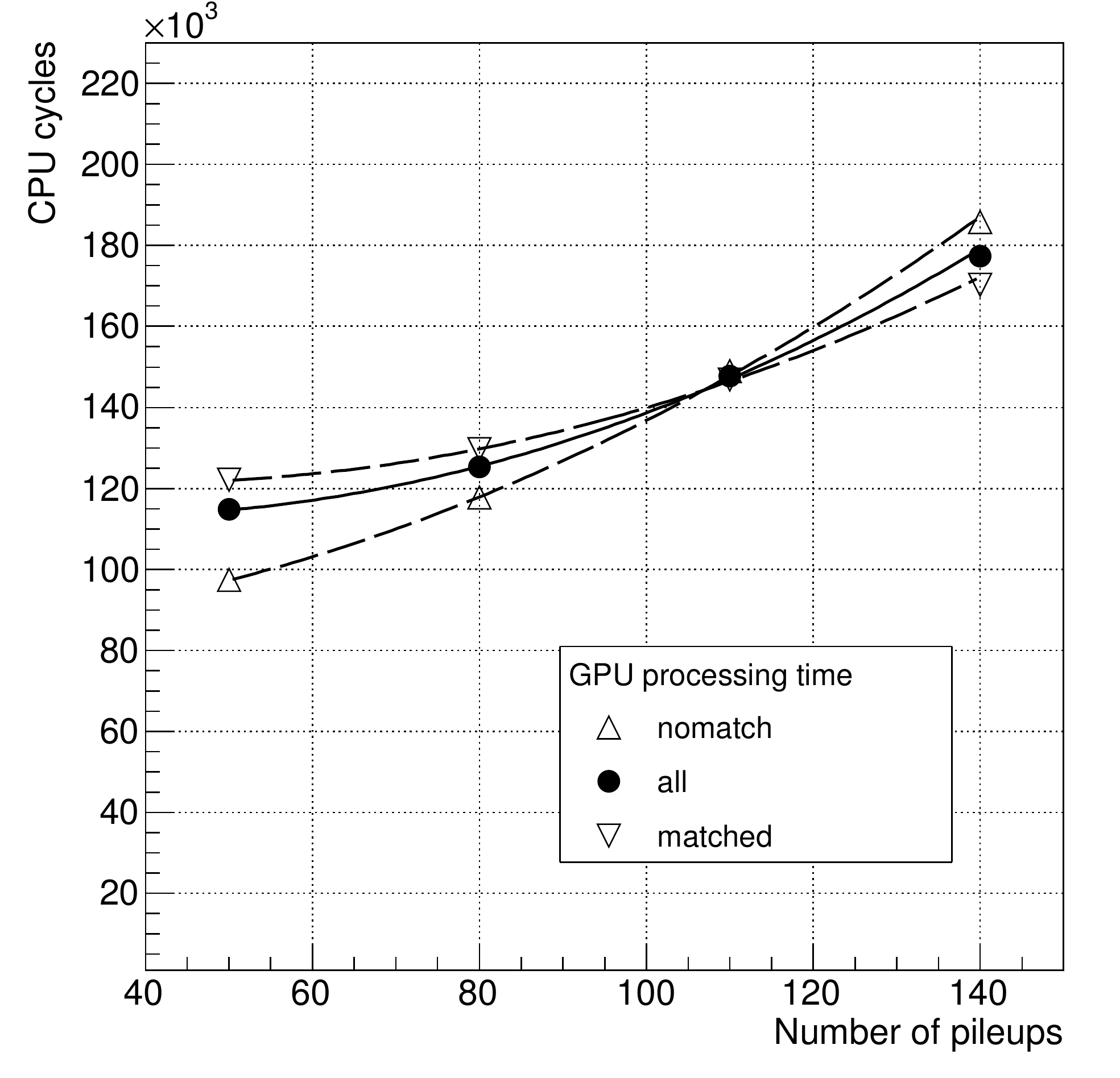} \\
\caption{The plot above shows the number of cycles to process the pixel track trigger algorithm described in Ref.~\cite{ref:pixtrk} as a function of the simulated data samples for the nVidia Tesla GPU described in the text.  Data-transfer times between host and GPU are not included in these results. The cycles are measured using the host Intel CPU's time stamp counter. The three different types of markers used have the same meaning as in Figure~\ref{fig:x86cycles}}
\label{fig:gpucycles}
\end{figure}

We also implemented the algorithm described in Ref.~\cite{ref:pixtrk} on an nVidia Tesla K40c GPU (745 MHz) using nVidia's CUDA programming environment.  In this case, the loops over 4-layer hit combinations were unrolled using parallel thread blocks where each thread block dealt with one hit combination from Layers 1 and 4.  Multiple threads in each block then dealt with Layer 2 and 3 hit combinations in parallel.  The number of processing cycles as a function of sample for the GPU are shown in Figure~\ref{fig:gpucycles}.  The processing cycles are measured using the host Intel CPU's time stamp counters.  The legend used in the graph is identical to that for the CPU with results shown for all clusters and separately for the two classes of clusters described above. The average GPU processing times are $34.8$, $38$, $44.8$, and $53.7$ $\mu$s, respectively, for the 50, 80, 110, and 140 pileup samples.
 
The GPU results do not show improvement over the CPU results, mainly because it is more complicated for the GPU to break out of a loop upon the first successful track match, or for it to skip to the next iteration of a loop through \emph{continue} statements. This makes the execution time dependent on the slowest thread.  The parallel capabilities of the GPU (with 2880 cores) are also not fully exploited by our test case where we have significantly reduced the combinatorics by considering only hits within the ROI. Furthermore, one must also take into account additional latencies associated with data transfers between the CPU and GPU, which contribute an additional $>91$K CPU cycles ($>27$ $\mu$s) to the total cycles (time). One place where the GPU does better, however, is on events in the tail of the execution time distributions of the CPU.  Such events have little influence on the spread of the GPU distributions because the GPU excels at dealing with problems that have more massive parallelism.
  
The algorithm described in Ref.~\cite{ref:pixtrk} used in the CPU/GPU comparisons above is functionally but not algorithmically equivalent to our automata-based track finder.  Using regular expressions to represent hit patterns, we also implemented an algorithmically equivalent NFA-based solution on the same GPU used above. We measure $\sim 10$ seconds to execute the algorithm for each cluster which is about 6 orders of magnitude longer than on the AP. 

\subsection{Note on CPU Processing Rate}\label{sec:cpurate}
For applications in which the processing latency is irrelevant and only the processing rate matters, it is interesting to see how many CPU cores it takes to process events in parallel in order to match the rate of the AP-based system.  Using the results for the sample overlaid with 140 pileup interactions and assuming that the $R-\phi$ and $R-Z$ views are done in parallel, the average single-event processing time (including the external processing time) for the AP-based system is $3.62$ $\mu$s, which is equivalent to an event rate of approximately $276$ kHz. Since the corresponding time on a single core of the Intel i7 CPU is $32.1$ $\mu$s, it would require about 9 CPU cores to match the processing rate of an AP-based system consisting of 490 AP chips.

\subsection{Associative Memories on an FPGA}
Unlike ASICs, FPGAs are programmable, off-the-shelf devices and should offer a fair comparison with the AP.  Reference~\cite{ref:pramposter} describes the development of a Pattern Recognition Associative Memory (PRAM) on modern FPGAs as part of Fermilab's tracking trigger R\&D program for the LHC experiments.  The PRAM, which is based on CAMs, is implemented using a mid-range Xilinx Kintex UltraScale KU040. Up to 4,000 patterns, for a detector consisting of 6 layers with 15-bit hit addresses, were stored, consuming 78\% of the FPGA resources~\cite{ref:camasics}.  The device was successfully operated at 250MHz with a fixed output latency of 7 clock cycles.   It is also important to note that, in this design, hits from all 6 detector layers are presented simultaneously to the PRAM on 6 parallel 15-bit input buses, in contrast to the single 8-bit bus of the AP.  The FPGA has an advantage over the AP in terms of maximum pattern capacity per chip and pattern finding speed. The advantage of the AP over the FPGA is in the ease with which it can programmed using the Micron AP SDK.

\section{Conclusion}
We have demonstrated a proof-of-concept use of the Micron Automata Processor in an electron track confirmation trigger for HEP.  Even the current, first version of this technology shows some promise for HEP trigger applications requiring low processing latencies.  In the AP's current form, CAM-based FPGA and ASIC implementations still surpass it in  terms of pattern storage capacity and processing performance.  It is clear that for specialized applications requiring the highest performance, such as the most demanding aspects of the lowest level trigger in a high-luminosity LHC environment, custom ASIC-based solutions may be the most sensible if not the only approach.  However, the availability of an off-the-shelf, dedicated pattern matching engine that is easy to program and suitable for HEP applications, provides a new alternative for situations (e.g. less demanding triggers and offline reconstruction) in which custom or even FPGA-based solutions would not have been considered previously.

Compared with other commodity off-the-shelf solutions like CPUs and GPUs, the AP requires a factor of over two orders of magnitude fewer hardware cycles to perform our sample track finding application. With a clock cycle of $7.5$ ns on the AP versus $0.3$ ns on the CPU, this lower hardware cycle count translates into a processing latency of $3.64$ $\mu$s on the AP that is $\sim 5 \times$ lower than that of the multithreaded CPU implementation. Lower processing latencies are crucial for the  online trigger application considered in this paper.  On the other hand, if we disregard processing latency and pay attention only to processing rate, it requires only 9 CPU cores to achieve the same processing rate as an AP-based system consisting of 490 AP chips.  Such requirements can be satisfied by a commodity server with dual 6-core CPUs.

When comparing the results, one must keep in mind that the Intel CPU used in this study is the 5th generation of a very mature technology.  Pre-production evaluation versions of the AP are only becoming available now at the time of this writing.  In our evaluation, we were also conservative in our choice of configurable parameters such as the size of the report vectors.  Choosing a smaller size can further reduce the times associated with the internal transfer and readout of these vectors.  Future versions of the AP may also incorporate larger alphabet sizes.  Doubling the current symbol recognition capability from 8-bits to 16-bits, for example, will cut the time to process 16-bit hit addresses in our track finding algorithm in half.
The possibility of such improvements, coupled with the results presented in this paper, suggest that this may be a promising technology worthy of more detailed consideration in real-world HEP pattern recognition applications.
Lastly, this signifies the first use of this interesting technology to the recognition of visual patterns.  It opens up a whole new realm which may even include image processing applications in fields like astronomy.

\section{Acknowledgements}
This paper is dedicated to the memory of Simon Kwan who worked tirelessly on the CMS pixel detector and upgrade project and who inspired our choice of the pixel-augmented electron trigger as a proof-of-concept application. We are grateful to David Christian, Aurore Savoy-Navarro, Chang-Seong Moon, Tiehui Ted Liu, Jin-Yuan Wu, Zijun Xu, and Ken Treptow for valuable discussions.  We thank the supportive staffs at the University of Virginia's Center for Automata Processing and Micron Technology for their technical assistance.  Fermilab is operated by Fermi Research Alliance, LLC under Contract No. De-AC02-07CH11359 with the United States Department of Energy.

\section*{References}

\bibliography{nim_aptrkfnd_refs}

\begin{thebibliography}{10}
\bibitem{ref:3dpats}
{Instead of this decomposition into 2 separate views, it would be possible, in
  principle, to have a single bank consisting of 3 dimensional patterns.
  Unfortunately, this is not practical because the number of patterns that
  needs to be stored makes it prohibitive in terms of resource usage on the
  automata processor.}
\bibitem{ref:latency}
{Available trigger latency at Level-1 is on the order of $\sim10$ $\mu$s for
  the era of the High-Luminosity LHC}.
\bibitem{ref:camasics}
{CAM-based ASIC implementations have achieved twice this density with 8K
  patterns per chip and the goal is to achieve 128K patterns per chip in the
  next generation of the chip~\cite{ref:amchip04,ref:atlasftktdr}}.
\end{thebibliography}

\end{document}